\documentclass[aps,twoside,twocolumn,groupedaddress,dvips,floatfix,a4paper]{revtex4}
\usepackage[intlimits]{amsmath}
\usepackage[]{graphicx}
\usepackage[USenglish]{babel}

\usepackage{amssymb}
\usepackage[dvips]{hyperref}
\usepackage[latin1]{inputenc}
\usepackage[squaren]{SIunits}
\usepackage{dcolumn}
\usepackage{mathptmx}
\usepackage{hyperref}

\begin{document}
\bibliographystyle{apsrev}

\newcommand{\nm}{\nano\meter}
\newcommand{\dose}{\centi\meter^{-2}}
\newcommand{\keV}{\kilo\electronvolt}
\newcommand{\eV}{\electronvolt}

\renewcommand{\arraystretch}{1.3}

\preprint{E-MRS 2002, Paper Reference: \textbf{S-IV.3}, REVISED VERSION}

\title[Nanocrystal Formation by Phase separation in ultra-thin gate oxides]{Nanocrystal
Formation in Si Implanted Thin SiO$_2$ Layers under the Influence of an Absorbing Interface}

\author{T. Müller}
\email[]{T.Mueller@fz-rossendorf.de}
\thanks{Tel. +49 351 260 3148; Fax. +49 351 260 3285}
\homepage[]{http://www.fz-rossendorf.de}
\author{K.-H. Heinig}
\author{W. Möller}
\affiliation{Forschungszentrum Rossendorf, Institut f\"{u}r Ionenstrahlphysik \\und
Materialforschung, PO-BOX 51 01 19, 01314 Dresden, Germany}

\date{\today}

\begin{abstract}
Kinetic 3D lattice Monte Carlo studies are presented on Si nanocrystal (NC) formation by phase
separation in 1 keV Si$^+$ implanted thin SiO$_2$ films. The simulation start from Si depth
profiles calculated using the dynamic, high-fluence binary collision code TRIDYN. From the initial
Si supersaturation, NCs are found to form either by nucleation, growth and Ostwald ripening at low
Si concentrations. Or at higher concentrations, non-spherical, elongated Si structures form by
spinodal decomposition, which spheroidize by interface minimization during longer annealing. In
both cases, the close SiO$_2$/Si interface is a strong sink for diffusing Si atoms. The NCs align
above a thin NC free oxide layer at the SiO$_2$/Si interface. Hence, the width of this zone denuded
of NCs has just the right thickness for NC charging by direct electron tunneling, which is crucial
for non-volatile memory applications. Moreover, the competition of Ostwald ripening and Si loss to
the interface leads at low Si concentrations (nucleation regime) to a constant width of the denuded
zone and a constant mean NC size over a long period of annealing.
\end{abstract}


\maketitle

\section{Introduction}
Nanocrystals (NCs) embedded in a matrix are known to form by phase separation of a supersaturated
solid solution. At low impurity concentrations, nucleation \cite{Hollomon:1953} and conservative
Ostwald ripening \cite{Ostwald:1897} will lead to an coarsening ensemble of precipitates during
annealing . According Lifshitz, Slyozov, and Wagner \cite{Lifshitz:1961,Wagner:1961}, the mean
particle size increases according a power law in time. At higher concentrations, phase separation
takes place by spinodal decomposition \cite{Cahn:1961}. A vanishing nucleation barrier
\cite{LeGoues:1984} implies that small initial concentration fluctuations amplify during the
annealing.  Non-spherical structures precipitate, which form a connected network of precipitates
above the percolation threshold \cite{Hayward:1987}. In the presence of an absorbing (reflecting)
interface phase separation is considerably altered. For spinodal decomposition, it is known that
the interface directs the wave vector of concentration fluctuations \cite{Binder:1998}. Hence, a
layered structure of precipitates forms close to the interface. Recent studies also revealed an
influence on nucleation, which can either be enhanced or depressed by the absorbing/reflecting
interface \cite{Brown:1994}.

\begin{figure}[tb]
 \includegraphics[width=0.95\columnwidth]{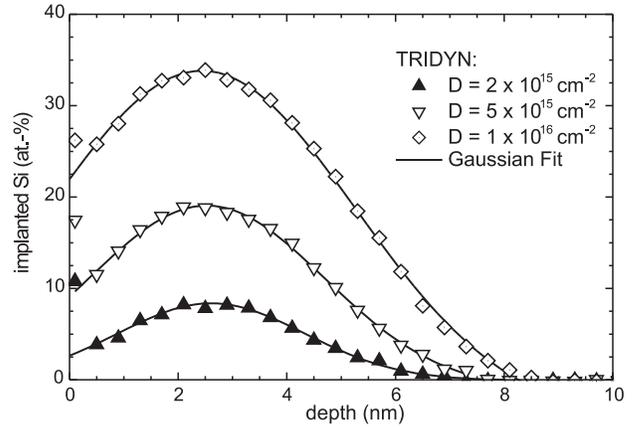}
 \caption{TRIDYN depth profiles for \unit{1}{\keV} Si$^+$ implantation into SiO$_2$ (denoted by
 open and bold symbols). The profiles were fitted to Gaussian distributions
 (Table~\ref{tab:gaussian}) shown as solid curves.} \label{fig:tridyn1}
 \newpage
\end{figure}
\begin{figure*}[t]
 \includegraphics[width=0.85\textwidth]{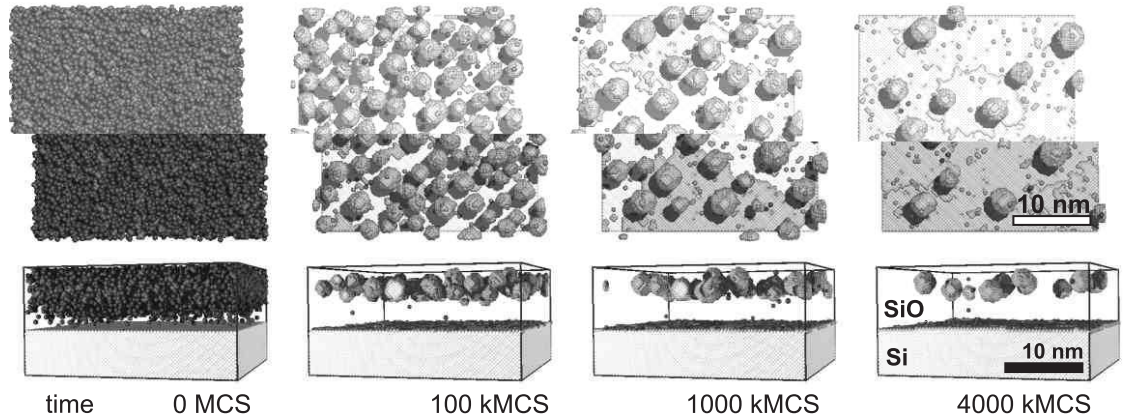}
 \caption{Snapshots of KMC simulations (top view and cross section) of phase separation in
 \unit{8}{\nm} thick SiO$_2$ on (001) Si implanted with Si. The initial Si depth profile was taken
 from Fig.~\ref{fig:tridyn1} for a fluence of \unit{2 \times 10^{15}}{\dose}.  Phase separation
 proceeds via nucleation and growth.} \label{fig:nucleation} \newpage
\end{figure*}

Phase separation in thin films is of particular interest for the fabrication of non-volatile
memories based on MOS transistors containing semiconductor NCs in their gate oxide
\cite{Tiwari:1996}. Using especially 1~keV Si$^+$ implantation into \unit{8}{\nm} thick SiO$_2$ on
(001)Si, NCs were formed a few nanometers above the Si/SiO$_2$ interface \cite{Normand:2001}, which
is in direct tunneling distance. Thus, these NCs can be charged by direct electron tunneling from
the Si, which is a prerequisite for a high endurance and low operation voltages of NC-memories
\cite{DeSalvo:2001}. Nevertheless, the influence of the absorbing SiO$_2$/Si interface on Si NC
formation is not well understood and requires further investigations. For that aim, kinetic 3D
lattice Monte Carlo (KMC) simulations of Si phase separation from SiO$_2$ were reported
recently\cite{Mueller:2002} and shall be discussed in more detail in this paper. As input data, Si
depth profiles for 1 keV Si$^+$ implantation into SiO$_2$ were taken from dynamic, high-fluence
binary collision simulations conducted with TRIDYN \cite{Moeller:1984}.
\section{Binary Collision Simulation of High Fluence Si$^+$ Implantation}
The TRIDYN depth profiles of implanted Si into of SiO$_2$ are shown in Fig.~\ref{fig:tridyn1} for
\unit{1}{\keV} energy. Similar to the widely applied binary collision simulation code TRIM
\cite{ZBL:1985}, TRIDYN is used to calculate ion range profiles in amorphous targets but includes
additionally dynamic target changes as induced by sputtering, swelling and ion beam mixing. The
input parameters being required include the displacement and surface binding energies of the target
atoms. For both, Si and O, the displacement energy was assumed to be \unit{8}{\eV}. The surface
binding energies of Si and O are assumed to vary linearly with the surface composition that they
balance the enthalpies of sublimation and decomposition of Si and SiO$_2$, respectively
\cite{Kelly:1980,TRIDYN:2001}.

The Si depth profile broadens with increasing fluence due to swelling, sputtering and  mixing
(Fig.~\ref{fig:tridyn1}. Additionally, preferential sputtering has lead to a Si enrichment at the
target surface. Nevertheless, the profile shape remains Gaussian in the fluence range considered
here. The parameters of Gaussian distributions fitted to the TRIDYN profiles of
Fig.~\ref{fig:tridyn1} are summarized in Table~\ref{tab:gaussian}, which also includes the
calculated target swelling.

\section{The kinetic Monte Carlo method}
\begin{table}[b]
\caption{Parameters of Gaussian distributions fitted to TRIDYN profiles of Fig.~\ref{fig:tridyn1}.
Additionally, the target swelling $d_{\text{swelling}}$ due to incorporated Si ions is given.}
\label{tab:gaussian}
\begin{ruledtabular}
\begin{tabular}{r|ccc}
  ion fluence (\unit{10^{15}}{\dose})& 2 & 5 & 10 \\ \hline
  projected range $R_p$ (\nm)        & 2.6 & 2.5 & 2.4 \\
  range straggling $\sigma$ (\nm)    & 1.8 & 2.3 & 3.0 \\
  peak concentration $c_{\text{max}}$ (at.-\%)  & 9   & 19  & 34  \\
  target swelling $d_{\text{swelling}}$ (\nm)   & 0.2 & 0.6 & 1.3 \\
\end{tabular}
\end{ruledtabular}
\end{table}

\begin{figure*}[tb]
 \includegraphics[width=0.85\textwidth]{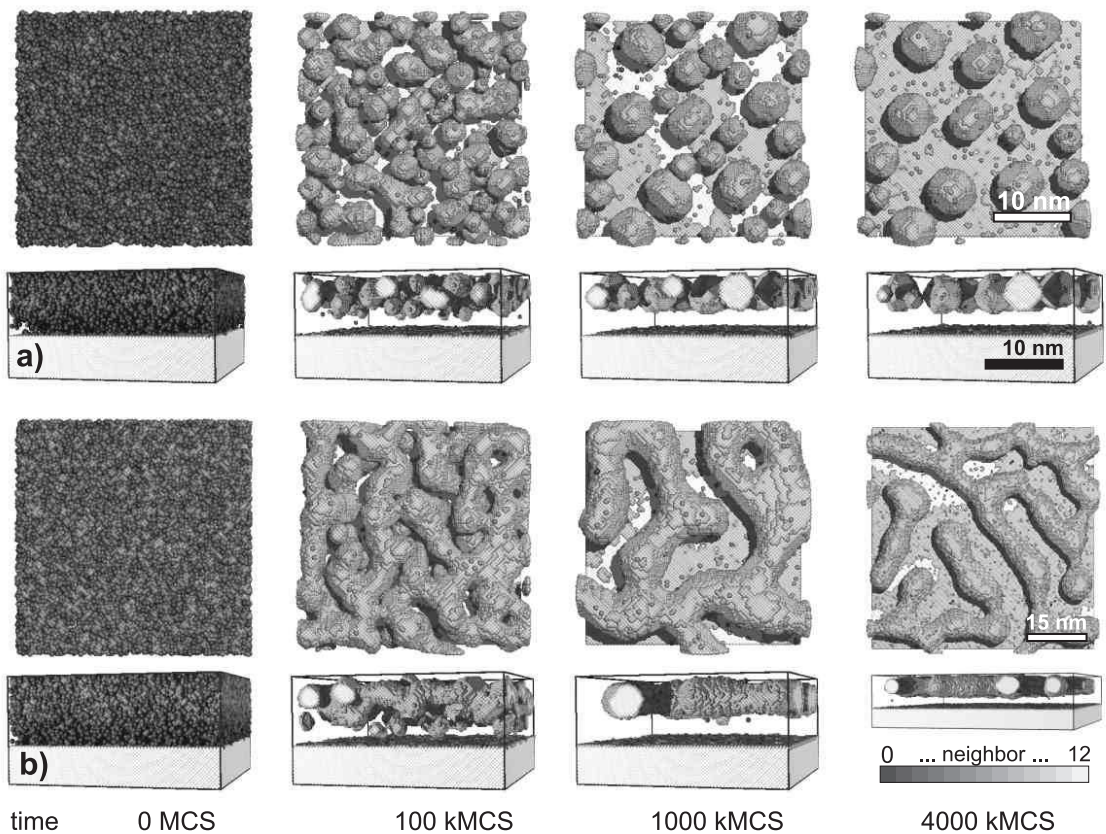}
 \caption{Snapshots of KMC simulations of phase separation
 in \unit{8}{\nm} thick SiO$_2$ on (001) Si implanted with Si. The initial depth profiles were taken
 according Fig.~\ref{fig:tridyn1} for fluences of \unit{5 \times 10^{15}}{\dose} (a) and \unit{1 \times 10^{16}}{\dose}
 (b). Phase separation proceeds via spinodal decomposition (a,b) and is accompanied by percolation at the
 highest fluence (b). The \unit{15}{\nm} scale applies only in the lower right corner.}
 \label{fig:spinodal}
\end{figure*}

The implanted Si separates from the SiO$_2$ matrix during subsequent annealing,  i.e.~precipitates
are formed. In general, this process of phase separation consists of a huge number of elementary
events (like bond breaking, diffusional jumps of atoms, chemical reactions etc.) that occur in
random sequence. Their simulation must take into account both energetic and statistical aspects,
which is here suitably conducted by kinetic 3D lattice Monte Carlo (KMC) simulations. The KMC
method applied is discussed elsewhere in detail \cite{Strobel:2001}, a short overview will be given
instead.

The kinetics of Si atoms is described in a solid host matrix (SiO$_2$), which is the background or
"system's vacuum". Thereby, an underlying fcc lattice has been assumed, which together with hcp is
the most isotropic lattice. Within this host dissolved Si diffuses and can form precipitates.
(Crystallites obey the same four-fold symmetry than in the diamond lattice. The lattice spacing
were chosen such that the correct atomic Si density is obtained.) Applying the classical lattice
gas model with attractive Si-Si interaction, the energetics is determined by the nearest-neighbor
(NN) Ising model. The kinetics of the system is governed by the Kawasaki particle exchange dynamics
using the Metropolis algorithm \cite{Metropolis:1953}. Statistically, each Si atom is allowed to
jump from an initial site $i$ to an empty neighboring site $f$ during one Monte Carlo step (MCS),
which is the time unit of the KMC simulation. The transition probability $P_{if}$ for one atom to
jump is given by
\begin{gather*}
    P_{if} =
    \begin{cases}
         \tau_0^{-1} e^{  - E_A \beta },    & n_f \ge n_i,\\
         \tau_0^{-1} e^{ - \left[E_A + (n_i-n_f)E_B\right] \beta},  & n_f < n_i,
    \end{cases}
\end{gather*}
where $\tau_0^{-1}$ denotes the attempt frequency, $n_{i,f}$ represents the number of NN bonds at
the two sites and $\beta=(kT)^{-1}$ has its usual meaning. $E_A$ is the diffusion barrier for Si in
SiO$_2$ and $E_B$ the Si-Si bond strength. In principle, the bond strength $E_B$ can be determined
from the solubility of Si in SiO$_2$ via the detailed balance of Si attachment/detachment at the
Si/SiO$_2$ interface. (It has to be taken into account that the coordination number in the fcc
lattice of the simulation is 12 instead of 4 in the case of the Si lattice.) In kinetic Monte Carlo
simulations it is convenient to renormalize the transition probability of the most probable event
to one. For $E_A \rightarrow 0$ each diffusional jump is allowed, which defines the temperature
dependent time scale of a Monte Carlo step (MCS), $\tau = \tau_0 \exp \{ E_A \beta \}$. Then, the
dimensionless transition probability reads as $\tilde{P}_{if} = min(1, \exp \left\{ - (n_i-n_f)E_B
\beta \right\})$. Energetically unfavored events ($n_i > n_f$) are allowed, nevertheless, to occur
by thermal activation according the Boltzmann statistics.

In principle, the scale of time and temperature of the simulation is given by the experimental
parameters of the system under consideration. However, the diffusivity and the solubility of Si in
SiO$_2$ are largely unknown, thus reduced units are applied for the time - the Monte Carlo step -
and for the temperature - the reduced temperature $E_B \beta$. Nevertheless, the path of systems's
evolution towards equilibrium predicted by KMC simulations may improve the process understanding
substantially.

\section{KMC Simulations of Phase Separation}
In Fig.~\ref{fig:nucleation} and \ref{fig:spinodal} snapshots of the KMC simulations of the phase
separation are shown for $E_B\beta=2$. Thereby, the TRIDYN profiles of Fig.~\ref{fig:tridyn1} were
used as initial Si distribution. The size of simulation volume is $56 \times 56 \times \Delta z$
$\nm^3$, where $\Delta z$ is \unit{8}{\nm} plus swelling due to implantation according
Tab.~\ref{tab:gaussian}. For the sake of clarity, only a quarter of the simulation cell is shown in
Fig.~\ref{fig:nucleation} and \ref{fig:spinodal} besides of the lower right corner in
Fig.~\ref{fig:spinodal}. Periodic boundary conditions were applied in the plane, while at the
surface a reflecting boundary conditions was applied. The simulation cell (SiO$_2$) borders on
fixed (001) layers of Si atoms accounting for Si attachment/detachment at the Si/SiO$_2$ interface,
which is an effective sink for diffusing Si atoms.
\begin{figure}[bt]
 \includegraphics[width=0.9\columnwidth]{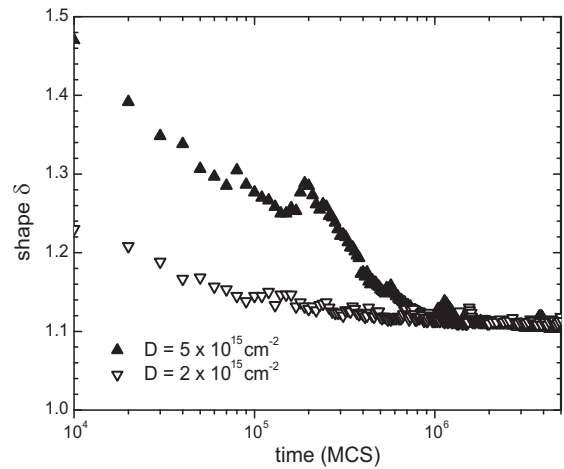}
 \caption{Evolution of mean NC shape measured by $\delta$ in the nucleation regime (\unit{2 \times
10^{15}}{\dose}) and for spinodal decomposition (\unit{5 \times 10^{15}}{\dose}). The highest
fluence (\unit{1 \times 10^{16}}{\dose}) is omitted due to percolation.}
 \label{fig:shape}
\end{figure}

Two regimes of phase separation are found by KMC simulations. A "nucleation and growth" regime is
observed for \unit{2 \times 10^{15}}{Si^+ \dose}. Si NCs nucleate and grow further at the expense
of Si supersaturation (Fig.~\ref{fig:nucleation}, 100 kMCS). Later on, NCs grow by Ostwald ripening
and finally dissolve due to Si loss to the SiO$_2$/Si interface, respectively. Above \unit{5 \times
10^{15}}{Si^+ \dose} a "spinodal decomposition" regime is identified, where due to the vanishing
nucleation barrier \cite{LeGoues:1984} non-spherical, elongated Si structures form
(Fig.~\ref{fig:spinodal} a, 100 kMCS). At even higher Si concentrations (\unit{1 \times
10^{16}}{Si^+ \dose}), above the percolation threshold, these structures become laterally connected
and extend over several tens of nanometers (Fig.~\ref{fig:spinodal} b, 100 kMCS).

As seen in Fig.~\ref{fig:nucleation} and \ref{fig:spinodal}, nucleation and spinodal decomposition
lead to different initial shapes of precipitates, which can be accessed by defining a shape
parameter $\delta$ \cite{Strobel:2001} (analogously to the moment of inertia),
\begin{gather*}
\delta = \frac{5}{3NR} \sum_{i=1}^{N} (\vec{r} - \vec{r}_i)^2.
\end{gather*}
It measures the mean squared distance of atoms from the center of mass of the NC relative to that
of a sphere with radius $R$, which assumes that all $N$ atoms within a NC form a sphere with
$R=\sqrt[3]{(3N)/(4\pi)}$. Thus, the shape parameter $\delta$ should tend to one for a spherical
NC. Non-spherical structures should obey $\delta > 1$. The time evolution of the shape parameter
$\delta$ is shown in Fig.~\ref{fig:shape}. In the nucleation regime, the NC shape deviates little
from the spherical form ($\delta \approx 1.2$) at the beginning, while it approaches quickly a
constant value of $\delta \approx 1.1$ during further annealing. (Small NC may have a $\delta$
slightly larger than one due to their facetted shape.) For spinodal decomposition $\delta$ the
formed structures are considerably non-spherical ($\delta >1.3$) initially. Below the percolation
threshold, non-spherical Si precipitates evolve into spherical NCs ($\delta \rightarrow 1$) due to
interface minimization, which can hardly be distinguished from NCs formed by nucleation (see
Fig.~\ref{fig:nucleation} and \ref{fig:spinodal} (a), 1000 kMCS).
\begin{figure}[t]
 \includegraphics[width=0.9\columnwidth]{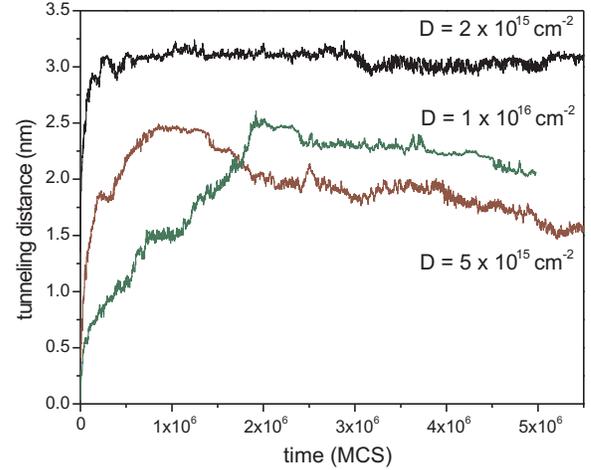}
 \caption{Width of the zone denuded of NCs between the SiO$_2$/Si interface and the NCs.}
 \label{fig:denuded-zone}
\end{figure}

\begin{figure}[tb]
 \includegraphics[width=\columnwidth]{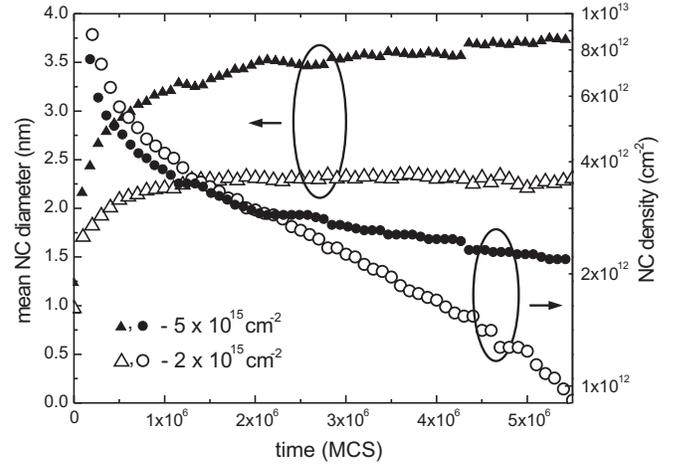}
 \caption{Evolution of the mean NC diameter and the NC density during annealing for fluences of
 \unit{2 \times 10^{15}}{\dose} and \unit{5 \times 10^{15}}{\dose}. The highest fluence (\unit{1
 \times 10^{16}}{\dose}) is omitted due to percolation.} \label{fig:size-density}
\end{figure}

The close SiO$_2$/Si interface acts as an effective sink for Si diffusing within the SiO$_2$ in
both regimes, which results in a zone denuded of NCs formed at the interface. However, a more
detailed consideration reveals differences between nucleation and growth and spinodal
decomposition. As shown in Fig.~\ref{fig:denuded-zone}, for the nucleation regime (\unit{2 \times
10^{15}}{\dose}) the width of the denuded zone is constant over a long period of annealing. A
similar behavior is observed for the mean NC size (Fig.~\ref{fig:size-density}). The competition of
NC dissolution due to Si loss to the interface and Ostwald ripening leads to a constant mean NC
diameter.

On the other hand, the mean precipitate size is larger in the spinodal decomposition regime
(Fig.~\ref{fig:spinodal}).  According the Gibbs-Thomson relation, the equilibrium Si solubility in
SiO$_2$ at the precipitate`s interface is lower than in the nucleation regime. Thus, the
diffusional flux of Si form the precipitates towards the Si/SiO$_2$ interface is greatly reduced.
Consequently, the NC diameter is still increasing. Within the annealing period shown in
Fig.~\ref{fig:size-density}, Ostwald ripening is more effective then the Si loss to the SiO$_2$
interface. Moreover, interface minimization of non-spherical precipitates leads to a narrowing of
the denuded zone (Fig.~\ref{fig:denuded-zone}).

In both regimes, NCs form at high density as shown in Fig.~\ref{fig:size-density} if the
percolation threshold is not reached. In the initial stage, the NC density is slightly higher in
the nucleation regime, while during longer annealing the NC dissolve faster than in the spinodal
regime.
\begin{figure}[tb]
 \includegraphics[width=\columnwidth]{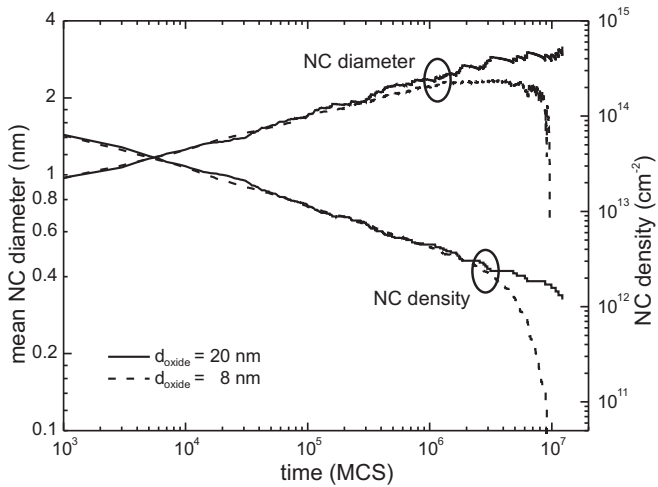}
 \caption{Comparison of mean NC size and density for phase separation in \unit{8}{\nm} and
 \unit{20}{\nm} thick SiO$_2$ layers implanted at a Si fluence of \unit{2 \time 10^{15}}{\dose}.}
 \label{fig:interface}
\end{figure}

For a fluence of \unit{2 \times 10^{15}}{Si^+ \dose} (nucleation regime), the evolution of mean NC
size and density is compared for phase separation in \unit{8}{\nm} and \unit{20}{\nm} thick SiO$_2$
(Fig.~\ref{fig:interface}). For both case, the same ion energy of 1 keV and, thus, the same initial
Si depth profiles were used. Up to an annealing time of \unit{10^{6}}{MCS}, the NC size and density
obeys the same power law for both oxide thickness (seen as straight line in the Log-Log plot of
Fig.~\ref{fig:interface}). The Si/SiO$_2$ interface does not influence the early NC evolution.
Later on ($> 1000$~kMCS), NCs start to dissolve in the 8 nm thick SiO$_2$ layer. Thus, the mean NC
size becomes constant and NCs dissolve faster than by Ostwald ripening, which is seen as kink in
the plot of the NC density after \unit{2 \times 10^{6}}{MCS}. The evolution of the NC size and
density is not affected by the far SiO$_2$/Si interface in the case of the thick (20 nm) SiO$_2$.

\section{Summary}

Binary collision simulations of high fluence 1 keV Si$^+$ implantation into 8 nm thick SiO$_2$ were
combined with kinetic 3D lattice Monte Carlo simulations of phase separation of Si from SiO$_2$.
Two major regimes of phase separation, nucleation and growth, and spinodal decomposition were
observed. At high concentrations, spinodal decomposition is accompanied by percolation. The regimes
of phase separation determines the initial shape of the precipitates. Nucleation and growth leads
to almost spherical NCs, while spinodal decomposition results in non-spherical Si structures, which
may spheroidize during long lasting annealing.

The performed process simulations predict that for non-volatile memory applications Si NCs should
be preferably synthesized in the nucleation regime. Then, NCs form at high density (\unit{>
10^{12}}{cm^{-2}}). The influence of the absorbing Si/SiO$_2$ interface leads to NC dissolution and
a zone denuded of NC forms at the interface. This denuded zone has just the right thickness of
\unit{2 \ldots 4}{\nm} in order to act a barrier for NC charging by direct electron tunneling from
the Si. Moreover in the nucleation regime, the width of this barrier is found to be constant over a
long period of annealing. A similar behavior holds for the mean NC size. For spinodal
decomposition, the situation is different. The width of the denuded zone and the mean NC size
varies during annealing (Fig.~\ref{fig:denuded-zone} and \ref{fig:size-density}). Above the
percolation threshold, the formed elongated structures are connected laterally, and, therefore,
would behave like a floating gate in a conventional MOS transistor. Charge brought to this network
of precipitates would spread and one oxide defect could lead to a complete discharging of the
floating gate, which could not occur in the case of laterally well-isolated Si NCs.


\begin{acknowledgments}
This work was sponsored by the European Community under the auspices of the GROWTH project.
GRD1-2000-25619. \end{acknowledgments}

\bibliography{Mueller2002-2}

\end{document}